# Discovery of an ultrasoft transient ROSAT AGN: WPVS007 *

**D. Grupe**[1], **K. Beuermann**[1,2], **K. Mannheim**[1], **H.-C. Thomas**[3], **H. H. Fink**[2], and **D. de Martino**[4]

[1] Universitäts-Sternwarte, Geismarlandstr. 11, D-37083 Göttingen, FRG
[2] MPI für extraterrestrische Physik, Giessenbachstr. 6, D-85740 Garching, FRG
[3] MPI für Astrophysik, Karl-Schwarzschild-Str. 1, D-85740 Garching, FRG
[4] ESA/VILSPA, PO Box 50727, E-28080 Madrid, Spain

**Abstract.** We have identified the ROSAT source RX J0039.2-5117 with the previously almost unknown 'narrow line' Seyfert 1 galaxy WPVS007 ($z = 0.028$). The X-ray source displays quite unusual properties for an AGN. It was bright and ultrasoft in the ROSAT All-Sky Survey (RASS) and found at a level lower by a factor of ~400 in PSPC count rate in 1993. The implied $0.1 - 2.4 \, \mathrm{keV}$ luminosity during the RASS was $\sim 10^{37}$ W. We discuss possible explanations for the extremely soft X-ray spectrum and the observed variability.

**Key words:** accretion, accretion disks – galaxies: active – galaxies: individual: WPVS007 – galaxies: nuclei – galaxies: Seyfert

## 1. Introduction

Among ROSAT's major achievements is the quantitative establishment of a pronounced soft X-ray excess over an underlying power law component in a large number of AGN. This excess may be part of a UV/soft X-ray bump and can be represented by a characteristic temperature kT in the range of $10 - 100 \, \mathrm{eV}$ (e.g. Walter & Fink, 1993). Ultrasoft X-ray emission in AGN was previously observed by Córdova et al. (1992) and evidence for pronounced soft X-ray variability in AGN was given by Piro et al. (1988), Yaqoob et al. (1994), Brandt et al. (1995), and Grupe et al. (1995). A strong soft excess is observed preferentially in Seyert 1 galaxies with narrow Balmer lines and strong FeII emission (Córdova et al. 1992, Puchnarewicz et al. 1992, Thomas et al. 1992, Grupe et al. 1994a,b, Boller et al. 1995). Accretion disc models have been favoured both for an explanation of the spectral bump and of the soft X-ray variability (e.g. Turner & Pounds 1989), although multi-frequency observations question this interpretation (Clavel et al. 1992).

In the course of our program of optical identifications of bright soft ROSAT sources, we found RX J0039.3–5117 to coincide with the practically unknown $V \simeq 15$, $z = 0.028$ emission-line object WPVS007, i.e., No. 7 in the list of Wamsteker et al. (1985), a Seyfert 1 galaxy with relatively narrow permitted lines and weak forbidden lines (Winkler et al. 1992). The X-ray source is peculiar because of its extreme softness and the large flux variation between the ROSAT All-Sky-Survey and a pointed ROSAT observation 3 years later, properties which are quite unusual for an AGN.

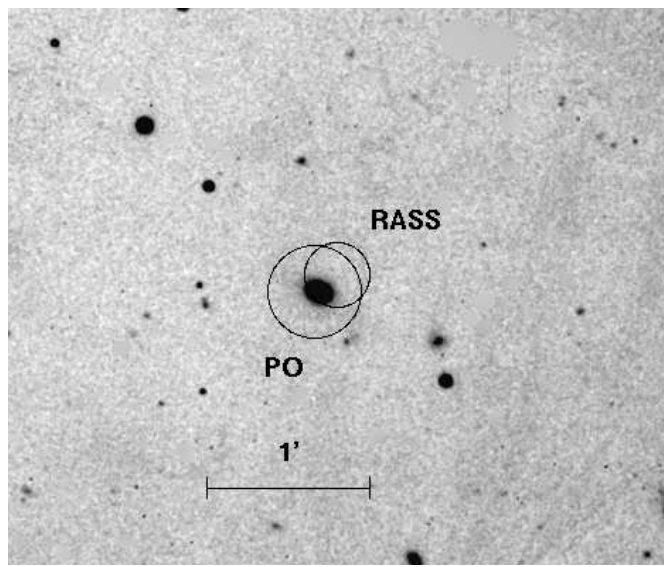

**Fig. 1.** ESO NTT V-band image of WPVS007. Also shown are the 95% confidence error circles of the X-ray positions from the RASS and ROSAT pointed observations. North is at the top and East to the left.

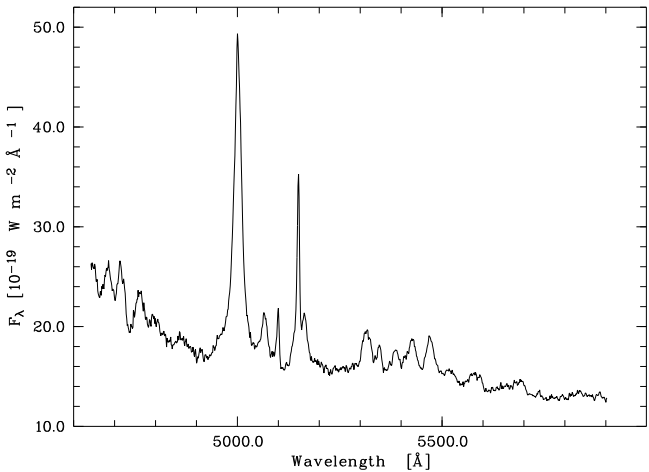

**Fig. 2.** Mean of four optical spectra of WPVS007 obtained with the 2.2m and 1.5m telescopes at La Silla between January 1993 and October 1994. The FWHM resolution is $\sim 5$ Å.

**Table 1.** Positions of WPVS007 and RX J0039.3–5117. The X-ray error radii refer to 95% confidence.

| Observation | $\alpha_{2000}$ | $\delta_{2000}$ | error radius (″) |
|---|---|---|---|
| Optical | $00^h 39^m 15\overset{s}{.}8$ | $-51° 17' 03''$ | 2 |
| RASS | $00^h 39^m 15\overset{s}{.}2$ | $-51° 16' 59''$ | 12 |
| ROSAT pointed | $00^h 39^m 16\overset{s}{.}5$ | $-51° 17' 09''$ | 17 |

## 2. Observations and results

### 2.1. Identification of the X-ray source

Fig. 1 shows a finding chart of WPVS007 from an ESO NTT V-band image taken on 9 February 1995 with the 95% confidence ($2\sigma$) X-ray error circles of the RASS and pointed ROSAT observations superimposed. These error radii (Tab. 1) are purely statistical and do not include possible systematic errors which could be of the order of 10″. The X-ray positions in Tab. 1 agree within the uncertainties with the optical position of WPVS007. No other object down to the plate limit was found within the X-ray error circles. The next object outside the error circles is a $V \simeq 22$ red object with some fuzzy emission nearby, located $\sim 20''$ SW of WPVS007 and $3.5\sigma$ and $2.3\sigma$ from the RASS and ROSAT pointed X-ray positions, respectively. This object is also visible on the ESO/SERC J and R plates. Its nature is unknown. Since nlSy1 galaxies are likely counterparts of high-galactic latitude soft X-ray sources, we consider the identification of the X-ray source with the much brighter Seyfert galaxy *in* the error circles as highly probable.

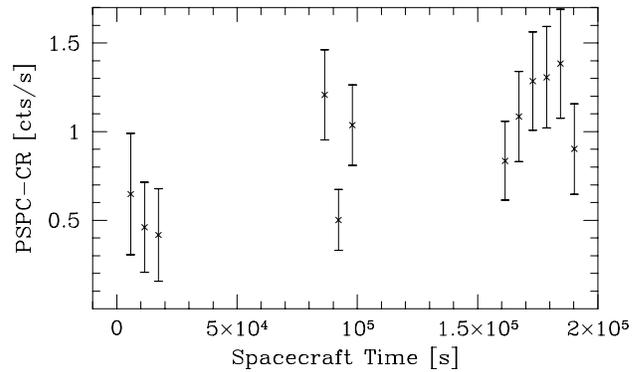

**Fig. 3.** X-ray light curve of RX J0039.2-5117 during the 2-day RASS observation.

**Table 2.** Mean $0.1 - 2.4$ keV PSPC count rates CR, hardness ratios HR1 (see text), and exposure times for the RASS and ROSAT pointed observations.

| Date | CR [cts s$^{-1}$] | HR1 | exposure [s] |
|---|---|---|---|
| 1990 Nov 10-12 | $0.98\pm0.08$ | $-0.94\pm0.06$ | 268 |
| 1993 Nov 11-13 | $0.0026\pm0.0014$ | $-0.86\pm0.14$ | 9691 |

### 2.2. Optical and UV observations

WPVS007 was discovered by Wamsteker et al. (1985) in a study of faint emission line objects and identified as a Seyfert 1 galaxy by Winkler et al. (1992). We observed WPVS007 four times between 1 January 1993 and 6 October 1994 with the ESO/MPI 2.2m and the ESO 1.5m telescopes at La Silla/Chile, covering the region around H$\beta$ each time and H$\alpha$ once. In the blue part, no spectral differences were seen and the mean is depicted in Fig. 2. The flux ratio of H$\alpha$ vs. H$\beta$ indicates little or no intrinsic reddening. With a FWHM(H$\beta$) $\simeq 1200$ km s$^{-1}$ and the strong FeII emission, we identify WPVS007 as a narrow line Seyfert 1 galaxy (nlSy1) in the classification scheme of Osterbrock & Pogge (1985).

Two IUE spectra of WPVS007 were taken on 5 September 1993 and 10 October 1994. They show Ly$\alpha$ in emission over a blue rather noisy continuum. Our combined optical/UV spectrum rises towards the UV as $F_\lambda \propto \lambda^{-2.3}$ (or $\nu F_\nu \propto \nu^{1.3}$, Fig. 5). WPVS007 is not listed as an IR or radio source in accessible catalogues.

### 2.3. ROSAT-PSPC observations

(1) *Count rates and hardness ratios:* Tab. 2 summarizes the results of the RASS and the pointed ROSAT observations performed with the Position Sensitive Proportional Counter (PSPC). Fig. 3 gives the light curve during the 2-day RASS observation. WPVS007 was a strong variable X-ray source during the RASS coverage in November 1990. It increased in count rate by a factor of ∼2 in two days. The mean hardness ratio[1] is HR1 = −0.94 ± 0.06 which implies that practically all photons have $h\nu < 0.4$ keV. This is the softest X-ray spectrum observed for a Seyfert galaxy and is rivalled in softness only by hot isolated white dwarfs, AM Herculis binaries, and certain other accreting white dwarfs (Beuermann et al. 1995).

**Table 3.** Best-fit X-ray spectral parameters for the RASS spectrum of WPVS007 assuming $N_H = N_{H,gal}$. The unabsorbed energy flux $F_x$ in the 0.1 − 2.4 keV band and the unabsorbed bolometric flux are in units of $10^{-15}$ W m$^{-2}$.

| Model | $\alpha$ | $kT$(eV) | $\chi^2/\nu$ | $F_x$ | $F_{bol}$ |
|---|---|---|---|---|---|
| *a) ROSAT All-Sky-Survey* | | | | | |
| Power law $\nu^\alpha$ | -7.3 | – | 5.7/11 | 860 | – |
| Blackbody | – | 20 | 6.3/11 | 170 | 450 |
| $\nu^\alpha \exp(-\frac{h\nu}{kT})$ | 0.3 | 41 | 9.0/12 | 79 | 390 |
| *b) ROSAT pointed observation* | | | | | |
| $\nu^\alpha \exp(-\frac{h\nu}{kT})$ | 0.3 | 18 | – | 4 | 180 |

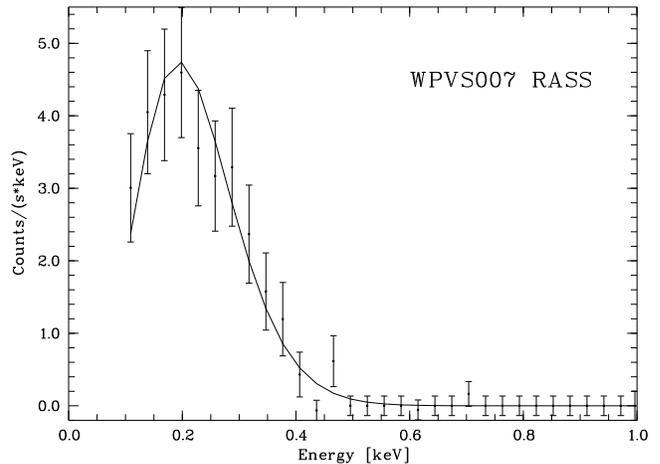

**Fig. 4.** RASS spectrum of WPVS007. The fit shown is a single blackbody with kT = 20 eV.

In the pointed observation three years later, WPVS007 had apparently 'turned off'. The count rate between the RASS and pointed observation decreased by a factor of ∼400 which is the largest factor reported so far for the X-ray variability in Seyfert galaxies. The X-ray spectral shape in the low state is still soft (Tab. 2). A comparable drop in ROSAT count rates (by a factor of ∼100) between RASS and later pointed observations was observed so far only in IC3599, another almost anonymous Seyfert galaxy (Brandt et al. 1995, Grupe et al. 1995).

(2) *PSPC-Spectrum:* Fig. 4 shows the mean PSPC spectrum from the RASS observation. It is extremely soft with all counts below 0.5 keV. A single power law $F_\nu \propto \nu^\alpha$ with energy spectral index $\alpha$ or a blackbody provide equally good fits. As there is no apparent absorption beyond that expected from galactic atomic hydrogen, the absorbing column density was fixed at $N_H = N_{H,gal} = 2.3\ 10^{20}$ H-atoms cm$^{-2}$ (Dickey & Lockman 1990). The power-law spectrum yields a best-fit slope of $\alpha \simeq -7$ and the blackbody a best-fit temperature $kT \simeq 20$ eV (Tab. 3). The very steep power law at soft X-ray energies exceeds the power-law extrapolation of the dereddened optical/UV spectrum at photon energies $h\nu < 0.16$ keV. At lower energies, the power law is unlikely to be representative of the true soft X-ray spectral flux. The blackbody may not be a valid representation either but naturally provides the required spectral turnover, staying well below the IUE flux at 1300Å.

As an alternative, we fitted the combined UV and X-ray spectrum by a model $F_\nu \propto \nu^\alpha \exp(-h\nu/kT)$ (Walter and Fink 1993) which yields a cut-off energy $kT \simeq 41$ eV. This spectrum (Fig. 5) stays below the blackbody and powerlaw fits to the RASS spectrum at the lower end of the PSPC range, i.e.

---
[1] The hardness ratio is defined as HR1 = *(hard-soft)/(hard+soft)*, where *hard* and *soft* refer to the count rates in the 0.4 − 2.4 and 0.1 − 0.4 keV energy ranges, respectively.

0.1 − 0.2 keV and may somewhat underestimate the soft X-ray/EUV flux at the time of the RASS (Tab. 3). The most simple explanation of the dramatically reduced PSPC count rate in the pointed ROSAT observation would be a reduction in the cut-off temperature of the exponential tail which would shift the soft component out of the PSPC window. A corresponding fit to the contemporaneous optical/UV/X-ray data for the pointed ROSAT observation is also shown in Fig. 5. The fluxes in Tab. 3 demonstrate that the dramatic reduction in PSPC count rate could be associated with a rather small reduction in bolometric flux, by a factor of ∼ 2, while the flux in the ROSAT band clearly drops by a much larger factor (Tab. 3). An important result is the complete absence of a detectable hard X-ray component within the ROSAT band.

The unabsorbed energy flux of soft X-rays in the ROSAT band 0.1 − 2.4 keV is $F_x \simeq 10^{-13}$ W m$^{-2}$. The correction factor to the bolometric (optical/UV/soft X-ray) flux is ∼4 (based on the non-contemporaneous UV spectrum and the $F_\nu \propto \nu^\alpha \exp(-h\nu/kT)$ model). For $H_o = 75$ km s$^{-1}$ Mpc$^{-1}$ and $q_o = 0.5$, the corresponding luminosities are $L_x \simeq 1.5\ 10^{37}$ and $L_{bol} \simeq 6\ 10^{37}$ W.

## 3. Discussion and Conclusions

The nlSy1 galaxy WPVS007 features unusual X-ray properties:

– the steepest AGN spectrum in the RASS,
– no detectable internal absorption on cold gas,
– no detectable hard X-ray tail,
– soft X-ray variability by two orders of magnitude over three years, and
– soft X-ray variability on a time scale of days.

In addition, there is no significant change in the optical spectrum before and after the RASS.

The fact that the long-term X-ray variability occurs in an almost anonymous galaxy raises the question whether WPVS007 is normally X-ray faint and just happened to be bright during the RASS. We have argued that this was the case in the likewise unknown AGN IC3599 which seems to have been in X-ray outburst at the time of the RASS, as judged from the decline in X-ray flux and in the Balmer and high-excitation Fe line fluxes after the RASS (Grupe et al. 1995, see also Brandt et al. 1995). For WPVS007, on the other hand, no long-term optical variability was observed and with only two X-ray points available, the nature of the long-term X-ray variability remains open.

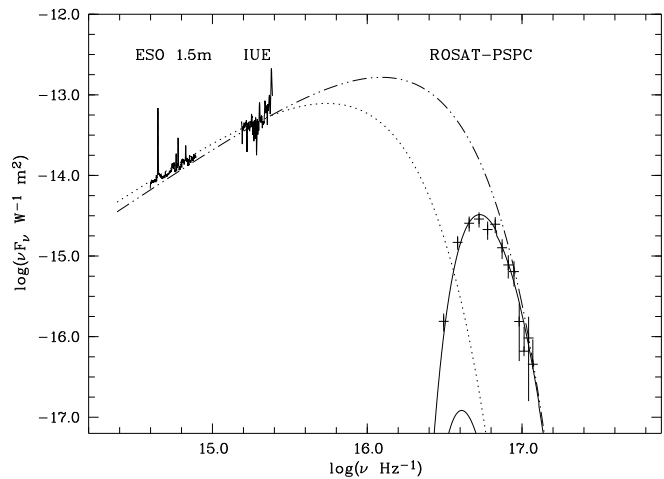

**Fig. 5.** Overall spectrum of WPVS007. Dereddened optical and UV spectra are contemporaneous to the pointed ROSAT observation, not to the RASS. *Dash-dotted curve:* $F_\nu \propto \nu^\alpha \exp(-h\nu/kT)$ respresentation to UV and RASS data, *dotted curve:* same to fit the low count rate from the pointed ROSAT observation (Tab.3). Note that the optical/UV spectral slope $\alpha = 0.3$ is remarkably close to the theoretical value for an extended accretion disk.

The bolometric luminosity of a few times $10^{37}$ W is not unusual for a nlSy 1 and the observed properties suggest a thermal origin of the X-rays. For disk accretion onto a black hole, the spectrum peaking in $\nu f_\nu$ at $\sim 10^{16}$ Hz favours a low mass of the black hole and a high (near Eddington) accretion rate during the RASS. The apparent turn-off could then be due to a variety of causes, including obscuration, a changing accretion rate, and changes in the nature of the radiative transfer. As demonstrated above, a small temperature change suffices to shift the Wien-tail of the RASS spectrum out of the PSPC band which supports an explanation in terms of a change in the inner-disk properties. Depending on the importance of the electron-scattering opacity in the inner disk, the corresponding change in bolometric luminosity need not be excessively large. Identifying the time scale of a few years with the viscous time scale in the disk would also favour a moderate black hole mass (Grupe et al. 1995).

A real surprise is the lack of a detectable hard X-ray flux. This lack fits into a scheme in which nlSy1s are Sy1s with weaker hard X-ray components rather that Sy1s with enhanced soft X-ray emission (cf. Boller et al. 1995). In such a scheme, however, the origin of the optical iron lines remains puzzling. Photoionization models associate the Fe II emission with the presence of hard X-rays. For WPVS007, this explanation fails completely except if the hard X-ray component is anisotropic (Ghisellini et al. 1991, Mannheim 1995). In that case, however, there should be detectable hard X-ray emission above $\sim 10$ keV due to Compton reflection which allows an observational test of this possibility. Another way out would be that the relativistic particles required to produce hard X rays in 'normal' Sy1 galaxies loose their energy by ionization of cold gas rather than by nonthermal radiation (Penston 1988). Finally, if the hard X-rays of Sy1s were of thermal origin (e.g. Haardt & Maraschi 1991), the optical depth of the hot comptonizing plasma would have to be negligible in WPVS007. In that case however, accretion should be sub-Eddington and the effective temperature of the accretion flow much lower than required to explain the observed soft X-rays.

To summarize, WPVS007 is an unusual nlSy1 galaxy which may become important for tests of the AGN paradigm and which requires further observations and detailed modelling.

*Acknowledgements.* We thank A. Goerdt and K. Reinsch for taking optical spectra of WPVS007, and E. Grebel for the V-image. We also thank B. Wills for useful comments and for providing the FeII template. We made use of the NASA Extragalactic Database (NED) operated by the JPL under contract with NASA and used IRAS data from the Infrared Processing and Analysis Center (IPAC). This research was supported by the DARA under grant 50 OR 92 10.